# Electronic and magnetic properties of $Ti_4O_7$ predicted by self-interaction corrected density functional theory

X. Zhong,[1] I. Rungger,[2] P. Zapol,[1] and O. Heinonen[1,3]

[1]Materials Science Division, Argonne National Laboratory, Lemont, Illinois 60439, USA
[2]School of Physics, AMBER and CRANN, Trinity College, Dublin 2, Ireland
[3]Center for Hierarchical Materials Design, Northwestern University, 2145 Sheridan Rd, Evanston, IL 60208

(February 13, 2015)

# Abstract

Understanding electronic properties of sub-stoichiometric phases of titanium oxide such as Magnéli phase $Ti_4O_7$ is crucial in designing and modeling resistive switching devices. Here we present our study on Magnéli phase $Ti_4O_7$ together with rutile $TiO_2$ and $Ti_2O_3$ using density functional theory methods with atomic-orbital-based self-interaction correction (ASIC). We predict a new antiferromagnetic ground state in the low temperature phase (or LT phase), and we explain energy difference with a competing antiferromagnetic state using a Heisenberg model. The predicted energy ordering of these states in the LT phase is calculated to be robust in a wide range of modeled isotropic strain. We have also investigated the dependence of the electronic structures of the Ti-O phases on stoichiometry. The splitting of titanium $t_{2g}$ orbitals is enhanced with increasing oxygen deficiency as Ti-O is reduced. The electronic properties of all these phases can be reasonably well described by applying ASIC with a 'standard' value for transition metal oxides of the empirical parameter α of 0.5 representing the magnitude of the applied self-interaction correction.

# 1. Introduction

Titanium dioxide (TiO$_2$) is a binary oxide that has been extensively investigated for its applications in photovoltaics[1], photocatalysis[2] and, recently, resistive switching[3-5]. A typical structure exhibiting resistive switching consists of a TiO$_2$ layer sandwiched between two metal electrodes. The resistance can be switched between two distinctive resistive states by the application of voltage pulses. It is now firmly established that oxygen vacancies play a crucial role[4, 6-8] in the switching mechanism. Stoichiometric TiO$_2$ can be easily reduced[9, 10,11] by external fields or by thermal means, leading to oxygen-deficient phases. When the concentration of the resulting oxygen vacancies is high enough, these phases may rearrange spontaneously to form ordered reduced structures, the so-called Magnéli phases (Ti$_n$O$_{2n-1}$). In fact, Magnéli phases such as Ti4O7 (n=4) have been identified experimentally[6, 9] in resistive switching devices, with evidence suggesting that filaments or regions of Magnéli phase Ti4O7 form conducting pathways. The formation of conductive filaments leads to a low-resistance state of the device, while the rupture of filaments leads to a high-resistance state.

Titanium oxide structures for use in applications such as photocatalysis and resistive switching are likely to be off-stoichiometric and contain a mixture of Ti-oxides. Therefore, understanding the behavior of Ti-O based structures that depends on the Ti oxidation states requires detailed knowledge of electronic properties of the different sub-stoichiometric phases of titanium oxide. From a theoretical and modeling point of view, it is known[7, 12, 13] that density functional theory (DFT) using the conventional local density approximation (LDA) or generalized gradient approximation (GGA) does not yield an accurate description of the electronic structure (e.g., band gap, or location of defect states within the band gap) of pristine and oxygen-deficient TiO$_2$. Recently, DFT methods beyond conventional DFT, specifically the LDA+U method[14] as well as hybrid functionals[15], have been used to study TiO$_2$, yielding in general satisfactory results.[7, 16, 17] For example, the experimental band gap of pristine TiO$_2$ is reproduced, and the relative stability of various vacancy charge states and the location of defect states in the band gap also agree well with experiments.

The crystalline Magnéli phase Ti$_4$O$_7$ has been shown experimentally to have a small energy gap (~0.1-0.2 eV) below 142 K (low temperature, or LT), while it is metallic at room temperature (high temperature, or HT)[18, 19]. In contrast, theoretical approaches using LDA+U[20, 21] or hybrid functionals[22] predict a much larger energy gap for the LT phase (1.5 eV and 0.75 eV, respectively) than the experimentally measured one. Furthermore, they predict contradictory electronic structures for the HT phase. Liborio *et al.* used the B3LYP hybrid functional[22] and obtained an insulating antiferromagnetic (AF) HT phase with a band gap of 0.4 eV. On the other hand, Weissmann et al. used LDA+U, with value of U (0.4 Ry) determined by comparing the relative energies of different Ti4O7 phases[21], and obtained a ferromagnetic (FM) metallic phase. Weissman et al.[21] obtained an energy difference between the predicted ferromagnetic phase and a semiconducting AF state that is only 0.01 eV per formula unit. This energy difference is very low and would result in a mixture of different states at room temperature. It is desirable to have a single approach that can properly describe at least some of the behavior that depends on the electronic structures of the Magnéli phase as well as those of $TiO_2$ and $Ti_2O_3$, considering that the Magnéli phase is positioned chemically between $TiO_2$ and $Ti_2O_3$. In fact, this is of great importance, especially in the context of modeling $TiO_2$-based resistive switching devices, since not only $TiO_2$ but also $Ti_2O_3$ may coexist with the Magnéli phase[10] when oxygen atoms are removed from $TiO_2$ as the material switches from a high to a low resistance state.

We have systematically modeled the electronic structures of the Magnéli phase $Ti_4O_7$ together with the end members of the Magnéli phase, i.e., $TiO_2$ and $Ti_2O_3$, with the aim of describing these structures and determining their ground states, especially the band gap and densities of states, within the same approach. These titanium oxide phases cover a broad range of electronic properties, from metal (HT-$Ti_4O_7$) to narrow-gap semiconductor[23, 24] (band gap~0.1-0.2 eV, LT-$Ti_4O_7$ and $Ti_2O_3$) and to insulator[11,24] (band gap~3 eV, rutile $TiO_2$). In our work, we used the atomic-orbital-based self-interaction correction (ASIC) scheme[25, 26]. The amount of self-interaction correction is controlled by a single parameter, $\alpha$, which lies between 0 and 1 (for $\alpha=1$ the full self-interaction correction is added, while for $\alpha=0$ no correction is added). The $\alpha$-parameter empirically describes the charge screening in the given chemical environment. In metals with very good screening $\alpha$ vanishes, while for highly ionic compounds with poor screening such as NaCl, a value of $\alpha$ close to unity reproduces the experimental band gap[25]. For III-V and II-VI semiconductors as well as transition-metal oxides, the appropriate value of $\alpha$ is shown to be typically around one half[25]. A central question regarding the ASIC methodology itself is how it compares to the other beyond semi-local DFT methods, and then whether a single value of $\alpha$ can be used to reasonably accurately describe a number of key properties of all the different titanium oxides.

In order to search for the low-energy phases of $Ti_4O_7$, we have investigated the total energies of all possible collinear spin configurations for both the LT and HT phases. The results for a fixed $\alpha$ value of 0.5, which is typically a reasonable value for transition metal oxides, are discussed in Sec. 3.1. A central result presented here is a new LT AF ground state, which is slightly lower in energy than the previously proposed[21] LT AF ground state, and we analyze the LT magnetic states and the energy difference between them using an effective Heisenberg model. In devices for resistive switching or photocatalysis, the presence of external electric fields or interfaces between different phases naturally induces structural deformations, i.e., strain, which may have a profound impact on the electronic properties of the materials involved. We have therefore looked into the effect of isotropic strain in Sec. 3.2 on the relative stability of different spin configurations for $Ti_4O_7$ to predict its ground state under strain. In Sec. 3.3 we discuss the variation of electronic structrue of Ti-O phases when stoichiometry changes.

## 2. Computational method

The DFT calculations are performed within the framework of the ASIC-LDA[25-27], as implemented in the SIESTA[28] electronic structure code. In a benchmark study[26] it was shown that the ASIC method is comparable to other advanced DFT methods (i.e., hybrid functionals and DFT+U) in delivering electronic structures, when applied to transition metal oxides such as MnO and NiO at equilibrium and under pressure. We use Troullier-Martins pseudopotentials[29] and double-ζ basis sets with polarization functions for Ti and O atoms. $Ti_4O_7$ has a triclinic structure with two pseudo-orthorhombic sublattices[30]. Each unit cell consists of two formula units with a total of 22 atoms (Fig. 1). The equilibrium lattice parameters (Table I) and atomic coordinates of $Ti_4O_7$ as well as $TiO_2$ and $Ti_2O_3$ are taken from experiments[31-33]. The SIESTA parameter 'kgrid_cutoff', which sets the density of sampling k-points in the Brillouin zone[34], is set to 30 Å to perform k-space integrations. This results in 486 points in the Brillouin zone for $Ti_4O_7$. The unit cell volumes in real space are 232.7 Å$^3$ and 232.3 Å$^3$ for LT-$Ti_4O_7$ and HT-$Ti_4O_7$, respectively.

TABLE I. Titanium oxides unit cell lattice parameters[31-33].

|  | a | b | c | α | β | γ |
|---|---|---|---|---|---|---|
| LT- $Ti_4O_7$ | 5.591 Å | 6.915 Å | 7.455 Å | 120.64° | 94.41° | 104.52° |
| HT- $Ti_4O_7$ | 5.593 Å | 6.899 Å | 7.441 Å | 120.56° | 94.46° | 104.35° |
| $TiO_2$ | 4.594 Å | 4.594 Å | 2.958 Å | 90° | 90° | 90° |
| $Ti_2O_3$ | 5.433 Å | 5.433 Å | 5.433 Å | 56.75° | 56.75° | 56.75° |

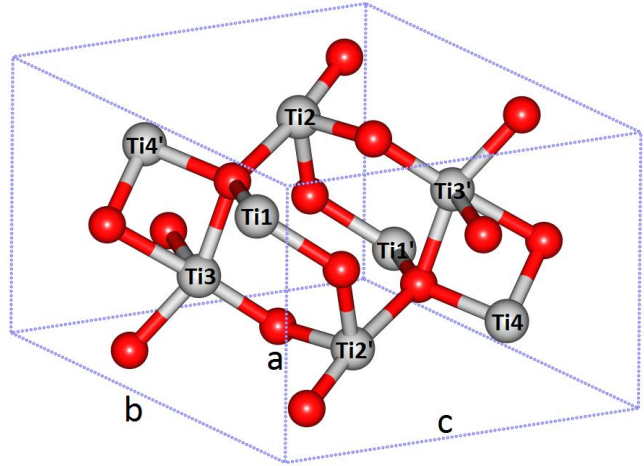

Figure 1. Atomic structure of $Ti_4O_7$ crystal unit cell. Grey spheres represent titanium and red spheres represent oxygen atoms. The numbering of Ti atoms is also shown.

## 3. Results and discussion

3.1 $Ti_4O_7$ spin configuration

We study $Ti_4O_7$ using a fixed value of the standard ASIC parameter $\alpha = 0.5$ at the two experimentally determined atomic structures given in Table I. It is well established[21, 22, 31, 35] that $Ti_4O_7$ exhibits charge localization at the metal-insulator transition. In the HT phase all Ti ions have uniform charges[22, 31] (+3.5), as evidenced by the similar average Ti-O distances

(2.01 - 2.02 Å) for all Ti ions. By comparing measured Ti-O distances and tabulated ionic radii these Ti ions were found[22] to have a charge of +3.5. Delocalization of the Ti 3d electrons leads to metallic properties of the HT phase. In contrast, in the LT phase the Ti ions can be classified into two groups: $Ti^{3+}$ (Ti1, Ti1′, Ti3 and Ti3′ in Fig. 1) and $Ti^{4+}$ (Ti2, Ti2′, Ti4 and Ti4′) ions. The corresponding average Ti-O distances are 2.04 Å for $Ti^{3+}$ ions, and 1.97/2.00 Å for $Ti^{4+}$ ions, respectively. Furthermore, long-range ordering of $Ti^{3+}$-$Ti^{3+}$ pairs leads to charge localization for the LT phase, which in turn causes semiconducting properties. Regarding the details of chemical bonding in $Ti_4O_7$, previous spin-resolved electronic structure calculations have yielded contradictory conclusions[21, 22], as discussed in the Introduction.

As shown in Fig. 1, eight Ti atoms within one unit cell correspond to two sublattices, with four Ti atoms in each sublattice. We use the same numbering (e.g., Ti1 and Ti1′) to denote the Ti atoms related by point group symmetry in the two sublattices, with the atoms in the second sublattice denoted by a prime sign. References 21 and 22 both considered only ferromagnetic coupling between the two sublattices, such that Ti1 and Ti1′ had the same magnetic moment, as did Ti2 and Ti2′ and so on. However, we note that the structural symmetry between the two sublattices is conserved by both ferromagnetic and antiferromagnetic coupling.

TABLE II. Total energies (per unit cell), magnetic moments (per unit cell), band gaps, averaged Ti-atom electric charge and Ti atomic spins estimated by Mulliken population analysis of different spin configurations as well the non-magnetic (NM) solutions of $Ti_4O_7$ at $\alpha = 0.5$. The energies shown are relative to the corresponding NM for both HT- and LT phases. Note that each unit cell consists of two formula units.

|  |  | HT |  |  |  | LT |  |  |  |  |
|---|---|---|---|---|---|---|---|---|---|---|
| Spin state |  | FM | AF1 | AF2 | NM | FM | AF1 | AF2 | AF3 | NM |
| Energy (eV) |  | -0.764 | -0.668 | -0.708 | 0 | -0.906 | -1.133 | -0.894 | -1.141 | 0 |
| Moment ($\mu_B$) |  | 2 | 0 | 0 | 0 | 2 | 0 | 0 | 0 | 0 |
| Band gap (eV) |  | 0 | 0.30 | 0.28 | 0 | 0.61 | 0.84 | 0.74 | 0.94 | 0.13 |
| Atomic spin ($\mu_B$) | Ti1 | +0.540 | +0.442 | -0.444 | 0 | +1.002 | +0.864 | +1.008 | +0.862 | 0 |
|  | Ti2 | +0.717 | +0.761 | +0.746 | 0 | +0.136 | +0.028 | -0.036 | -0.004 | 0 |
|  | Ti3 | +0.554 | +0.370 | -0.456 | 0 | +0.890 | -0.861 | +0.872 | -0.861 | 0 |
|  | Ti4 | +0.435 | +0.036 | +0.033 | 0 | +0.132 | -0.043 | -0.002 | +0.028 | 0 |
|  | Ti1′ | +0.540 | -0.442 | +0.444 | 0 | +1.002 | +0.864 | -1.008 | -0.862 | 0 |
|  | Ti2′ | +0.717 | -0.761 | -0.746 | 0 | +0.136 | +0.028 | +0.036 | +0.004 | 0 |
|  | Ti3′ | +0.554 | -0.370 | +0.456 | 0 | +0.890 | -0.861 | -0.872 | +0.861 | 0 |
|  | Ti4′ | +0.435 | -0.036 | -0.033 | 0 | +0.132 | -0.043 | +0.002 | -0.028 | 0 |

In order to determine the ground state spin configuration, we explore all possible spin configurations that respect the structural symmetry. Explicitly, we consider all 8=2⁴/2 (flipping all atomic moments simultaneously returns the same state) possible spin configurations across 4 Ti atoms within each sublattice (++++, +++-, ++--, …, +---), as well as both FM and AF coupling between the two sublattices. Thus, we consider all 16 possible collinear spin states within a unit cell. After the total energies converge self-consistently, we find that some initial spin configurations are transformed to other states, indicating that they do not correspond to a metastable local minimum in energy. In all, we find three stable spin-polarized solutions for the HT-phase (one FM state and two AF states), and four stable solutions for the LT-phase (one FM state and three AF states) (Table II). For the LT phase, we predict two competing ground states: the AF1 and AF3 states. Both states are antiferromagnetic with similar band gaps, while AF3 is 8 meV/unit cell lower in energy than AF1. It is to be noted that each unit cell consists of two formula units of $Ti_4O_7$, see Fig. 1. The AF1 state is very similar to the ground state obtained by either the B3LYP hybrid functional[22] or the LDA+U approach[21]. Both the AF1 and the AF3 states exhibit AF coupling between $Ti^{3+}$ ions within each sublattice. The difference lies in how the two sublattices are coupled to each other: in the AF1 state two sublattices are ferromagnetically coupled, while in the AF3 state they are antiferromagnetically coupled. In the higher-energy AF2 state on the other hand two $Ti^{3+}$ ions are coupled ferromagnetically, while the two sublattices are antiferromagnetically coupled. The magnetic moments for the $Ti^{4+}$ ions and all oxygen ions are very small, since they do not have localized partially filled orbitals.

The energy difference between the LT magnetic states can be attributed to (magnetic) exchange interactions, as we will now show. First, the four magnetic states can be classified into two groups by energy, the FM-AF2 group (Group I), and the AF1-AF3 group (Group II). Within each group, the Mulliken populations on each $Ti^{3+}$ ion are equal to within less than 0.002 $e^-$ (not shown), and spins on each $Ti^{3+}$ ion are also approximately equal (Table II).

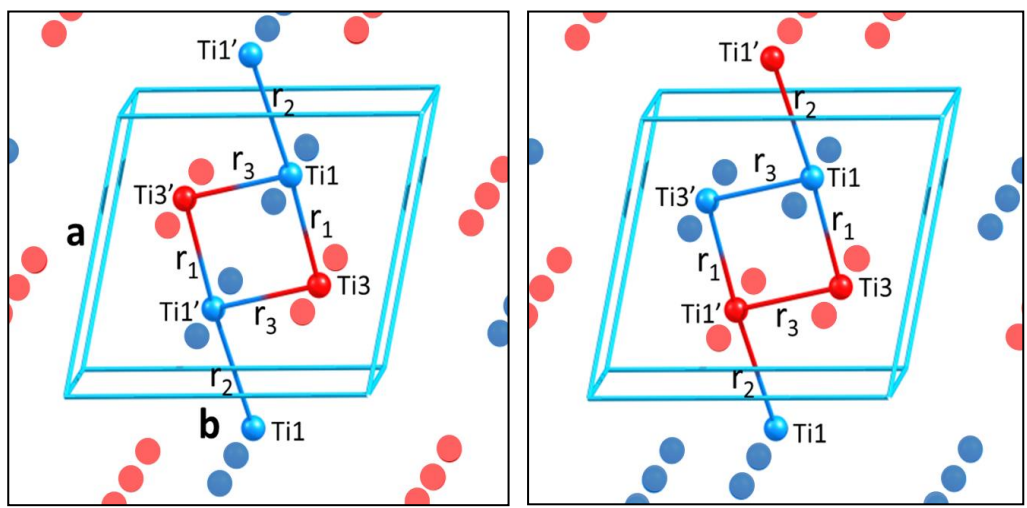

Figure 2. The spin topology of the AF1 state (left panel) and the AF3 state (right panel) in LT-$Ti_4O_7$. The blue and red dots represent spin up and spin down, respectively. Atomics spins of the atoms in the center unit cell and those connected to the cell are highlighted by brighter colors.

This indicates that each $Ti^{3+}$ ion has very similar local chemical environment in both states in each group, and that, if we neglect the spin polarization, both states in each group should yield degenerate energies (same direct Coulomb energy). In Fig. 2 we depict the spin structures of AF1 state and AF3 state of the LT phase in real space to better understand the difference between these two states. In both spin states the atomic spins of $Ti^{4+}$ ions are less than 0.05 $\mu_B$ (Table II), and in the FM and AF2 states the spin on these ions is less than 0.14 $\mu_B$. Therefore, we omitted these atoms in Fig. 2 for better visualization, and we show only the $Ti^{3+}$ ions (Ti1, Ti1′, Ti3 and Ti3′). There are three nearest neighbor distances between the $Ti^{3+}$ ions: $r_1$, $r_2$ and $r_3$, as indicated in Fig. 2. The lengths of these distances are $r_1 = 2.802$ Å, $r_2 = 3.133$ Å and $r_3 = 3.159$ Å, respectively. The next shortest interatomic distance is about 3.3 Å. Since exchange interactions in general decay quickly with interatomic distance, we include only interactions between atoms with the three smallest interatomic distances. We analyze the energy difference within the manifold of LT states using a Heisenberg model, as will be shown in the following. Based on the discussion above, we write a Hamiltonian for the LT-manifold that separates the interactions into exchange interactions and direct Coulomb and correlation interactions as

$$H = -J_1(\vec{s}_1 \times \vec{s}_3 + \vec{s}_{1'} \times \vec{s}_{3'}) - J_2 \vec{s}_1 \times \vec{s}_{1'} - J_3(\vec{s}_1 \times \vec{s}_{3'} + \vec{s}_{1'} \times \vec{s}_3) + H_c. \qquad (1)$$

Here, $\vec{s}_1, \vec{s}_2,$ and $\vec{s}_3$ ($\vec{s}_{1'}, \vec{s}_{2'},$ and $\vec{s}_{3'}$) are (dimensionless) spin operators on Ti1, Ti2, and Ti3 (Ti1', Ti2', and Ti3') and we use a spin quantization axis in which the "up" and "down" components are diagonal, and $H_c$ contains the direct Coulomb energy, correlation energy, etc.; the coupling constants $J_1$, $J_2$ and $J_3$ correspond to interactions between pairs of atoms separated by $r_1$, $r_2$ and $r_3$, respectively (Fig. 2). We can evaluate the energies of the LT states using Eq. (1) assuming that the expectation value of $H_c$ yields the same constant within the LT manifold of states. For the AF1 and AF3 states we obtain the energies

$$E_{AF1} = 2s_0^2 J_1 - s_0^2 J_2 + 2s_0^2 J_3 + C, \qquad (2)$$

and

$$E_{AF3} = 2s_0^2 J_1 + s_0^2 J_2 - 2s_0^2 J_3 + C, \qquad (3)$$

respectively. In Eq. (1) $C$ is the expectation value of $H_c$, and $s_0$ are the expectation values of the spin operators on Ti$i$ and Ti$i'$, $i=1,2,3$ in AF1 and AF3, which we take to be equal ($s_0 \approx 0.86$, see Table II). The energy difference between AF1 and AF3 is then obtained as

$\Delta_2 = E_{AF3} - E_{AF1} = 2s_0^2 J_2 - 4s_0^2 J_3$. Because $\Delta_2$ is negative from our ASIC calculations (Fig. 3), $J_2 - 2J_3$ is also negative. For the FM and AF2 states, we can similarly write

$\Delta_1 = E_{AF2} - E_{FM} = 4s_1(J_2 s_1 + J_3 s_3)$, where $s_1 \approx 1.0$ and $s_3 \approx 0.89$ (Table II), and we are assuming that the coupling constants $J_1$, $J_2$ and $J_3$ are the same in the two groups. We can evaluate $J_2$ and $J_3$ from the energy differences $\Delta_1$ and $\Delta_2$ to obtain $J_2 \approx -0.6$ meV and $J_3 \approx 3$ meV. Finally, we can use the energy difference between the AF3 and FM states, or between the AF3 and AF2 states, to estimate $J_1$, which yields $J_1 = -72$ meV or $J_1 = -78$ meV, respectively. The uncertainty in the estimate for $J_1$ can be attributed to the fact we ignored the small spins on the $Ti^{4+}$ ions in FM and AF2. This simple analysis demonstrates that the energy differences between the LT magnetic states can be explained by near-neighbor exchange coupling, and that the overall energy scale between the groups is set by $J_1$, while the energy differences within the magnetic states in each group are controlled by $J_2$ and $J_3$. This is also consistent with the $r_1$ distance considerably shorter than $r_2$ and $r_3$.

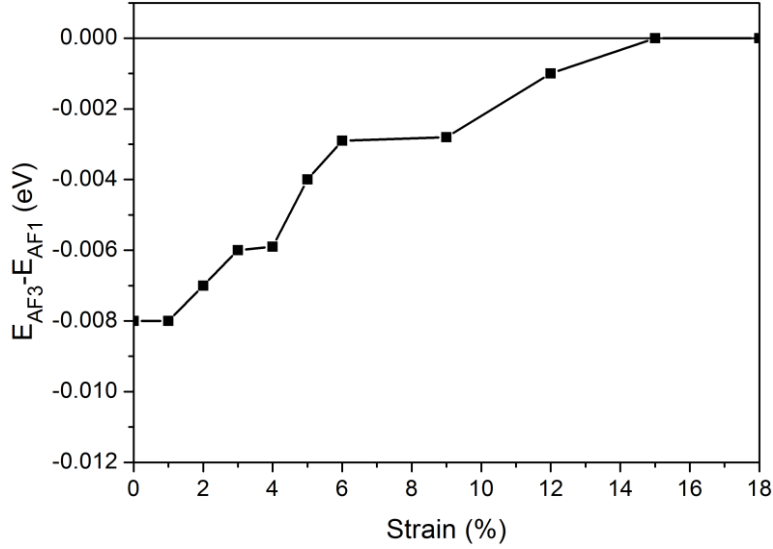

Figure 3. The evolution of energy difference $\Delta_2$ between the AF1 state and the AF3 state of LT-$Ti_4O_7$ as a function of positive isotropic strain.

In applying the Heisenberg model we have made the assumption that the exchange coupling at large distances is negligible. To justify this assumption we model the effect of positive isotropic strain (applied tensile stress) on the energy difference between the AF1 and AF3 states of the LT phase (Fig. 3) . The isotropic strain is modeled by expanding the lattice constants and all atomic coordinates proportionally. As a result, $r_1$, $r_2$, and $r_3$ all expand proportionally. Fig. 3 shows that $\Delta_2$ remains negative under the expansion up to +15% strain, beyond which the energies of both states become identical. Under the assumption that the energies are dominated by exchange interactions, this indicates that the exchange interactions vanish at 15% strain and beyond. At +15% strain the maximum relevant distance is $r_3$ = 3.633 Å, so we can deduce that exchange interactions become negligible when interatomic distances between Ti atoms are greater than 3.63 Å.

Because the predicted relative energy of the AF3 state is only 8 meV per unit cell lower with respect to the AF1 state using the DFT-ASIC method, we have also applied a hybrid functional, HSE06[36, 37], as implemented in the all-electron DFT code, FHI-aims[38], in order to verify that our results for AF1 and AF3 are robust with respect to the calculation method. In the HSE06 functional, 25% of the exact Hartree-Fock (HF) exchange energy is split into a short-range part with a screening parameter of 0.11 bohr$^{-1}$, and a PBE[39] GGA-like functional for the long-range exchange. In spite of the fact that the applied ASIC method makes use of pseudopotentials, while the adopted hybrid functional is implemented in an all-electron code, both methods yield quite similar energy difference between AF1 and AF3. The latter method predicts an energy difference of 10 meV per unit cell, with AF3 also being the state with the lower energy.

The density of states (DOS) for the AF1 and AF3 states are shown in Figure 4. The ASIC and hybrid functional methods give results with very similar DOS features for both states, except for the fundamental gap: the energy gaps predicted by HSE06 are about 0.5 eV larger than those by ASIC. In the AF1 state two sublattices of $Ti_4O_7$ are ferromagnetically coupled, while Ti1 and Ti3 (or Ti1″ and Ti3′) are antiferromagnetically coupled. Although the total magnetization is zero, spin up DOS and spin down DOS are not degenerate for AF1. On the other hand, the two spin components are degenerate in our LT-AF3 ground state, as a result of AF coupling between two sublattices.

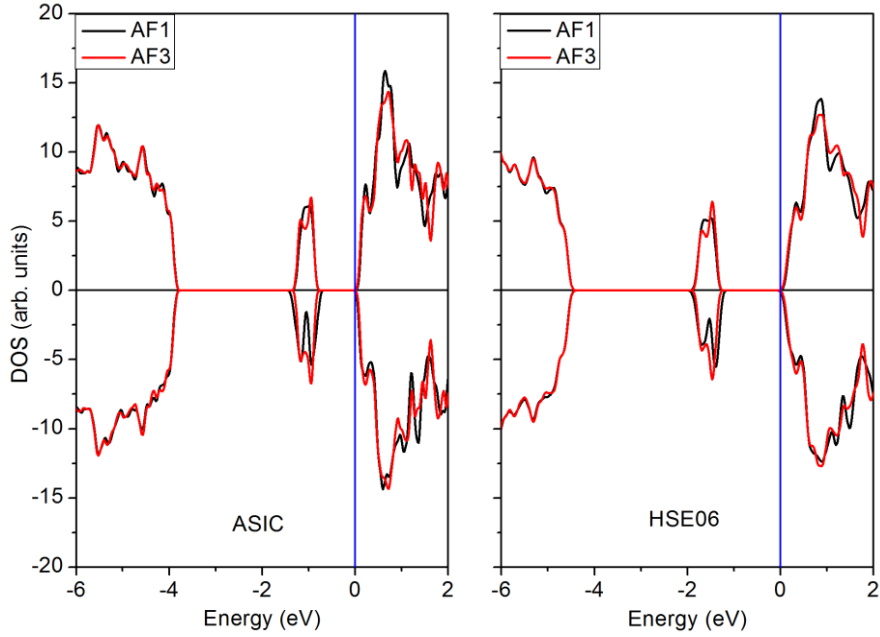

Figure 4. Density of states of the competing AF1 and AF3 states, calculated by both the ASIC method (left panel) in SIESTA and the HSE06 hybrid functional in FHI-aims (right panel). The black and red lines represent spin up and spin down, respectively. The Fermi level (energy zero) is aligned with the conduction band minimum (CBM).

For the HT phase, $\alpha = 0.5$ results in a metallic ferromagnetic ground state. We use Mulliken population analysis to estimate atomic spins presented in Table II; some oxygen atoms have small but non-zero opposite spins. All Ti atoms have similar atomic spins (0.4 - 0.7$\mu_B$), reflecting similar chemical environments for them. The total magnetic moment is 2 $\mu_B$ per formula unit, which corresponds to a full spin polarization of localized 3d electrons of all Ti ions, with an average formal charge of 3.5. Our result is similar to the FM state predicted by Weissmann[21] (LDA+U), which, however, was not the lowest-energy state in that work. We note that we also obtain two AF states, AF1 and AF2 (Table II), relatively close in energy to the FM state with energy differences of only 0.1 and 0.05 eV per unit cell, respectively, relative to the FM state. Magnetic susceptibility measurements suggest that $Ti_4O_7$ is Pauli-paramagnetic[18] at room temperature, but with a large susceptibility. Our results, as well as those of Weissmann and Weht[21], show a complex energy landscape with several competing FM and AF states. This suggests that $Ti_4O_7$ at room temperature may contain a mixture of FM and AF (as well as non-magnetic) states, the combined effect of which is to yield a large, but paramagnetic, susceptibility. For the purpose of electronic conduction, or other properties that depend primarily on the bandgap, the ASIC method with α=0.5 correctly yields a metallic state for $Ti_4O_7$. Therefore the ASIC method is applicable for studying resistive switching in Ti-oxide based devices as it correctly describes the bandgap evolution across the Ti-oxide series from $TiO_2$ to $Ti_4O_7$.

3.2 Evolution of spin states with applied strain.

All calculations presented in section 3.1 are based on experimental structures. For the transition metal oxides MnO and NiO, it was found that LDA+ASIC underestimates the experimental lattice constant[26] by ~2%. It was also shown in the same work that the electronic spin polarization depends critically on the lattice constant, i.e., strain. Table II shows that the total energy differences between some spin states of $Ti_4O_7$ are small (< 0.01 eV), which may lead to strain-induced magnetic phase transition[26]. This is an issue that needs to be considered, because strain fields are expected to be present in resistive switching devices due to the presence of, e.g., external electric fields or interfaces.

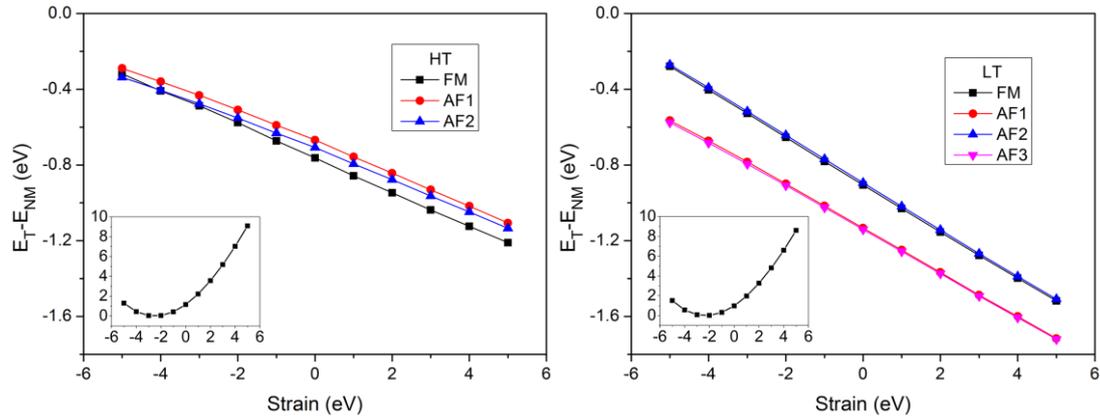

Figure 5. Differences between the ASIC ($\alpha = 0.5$) total energy, $E_T$, of the various spin configurations of $Ti_4O_7$ and the total energy of the nonmagnetic state, $E_{NM}$, as a function of isotropic strain. Positive strain values correspond to tensile stress while negative strain values correspond to compressive stress. Total energies of the non-magnetic states, $E_{NM}$, vs. applied strain are shown in the inserts.

The evolution of total energies of different spin configurations (relative to the respective non-magnetic LT and HT states) as a function of isotropic strain is shown in Fig. 5. We model isotropic strain by varying lattice constants and all atomic coordinates proportionally. For both the HT phase and LT phase the equilibrium structures modeled by ASIC underestimate experimental lattice constants by about 2%, which is consistent with previous ASIC benchmark work for other oxides[25]. It is clear from the figure that spin polarization always lowers the total energy compared to non-polarized states. While the HT phase appears to show an FM-AF2 magnetic transition at around -4% strain, there is no transition for the LT phase in the full modeled strain range. We note that the ASIC method applied here is not variational with respect to atomic coordinates[40] and therefore the total energy vs. strain may not be completely accurate. A variational pseudo-self-interaction-corrected density functional approach[40] (VPSIC), which is beyond the scope of this work, can potentially give more accurate total energies. VPSIC predicts essentially the same results as ASIC does for electronic structure calculations with fixed atomic positions[40] We expect the ASIC predicted FM to AF2 transition for the HT phase to be robust, although the exact value of the strain at which it occurs may be slightly different in magnitude. We note also that the energy difference between the LT-AF1 and LT-AF3 states at the experimental structure is very similar for the ASIC and hybrid functional calculations, which lends confidence that calculated relative energy for the different spin states by ASIC is robust.

3.3 Electronic structure vs. stoichiometry

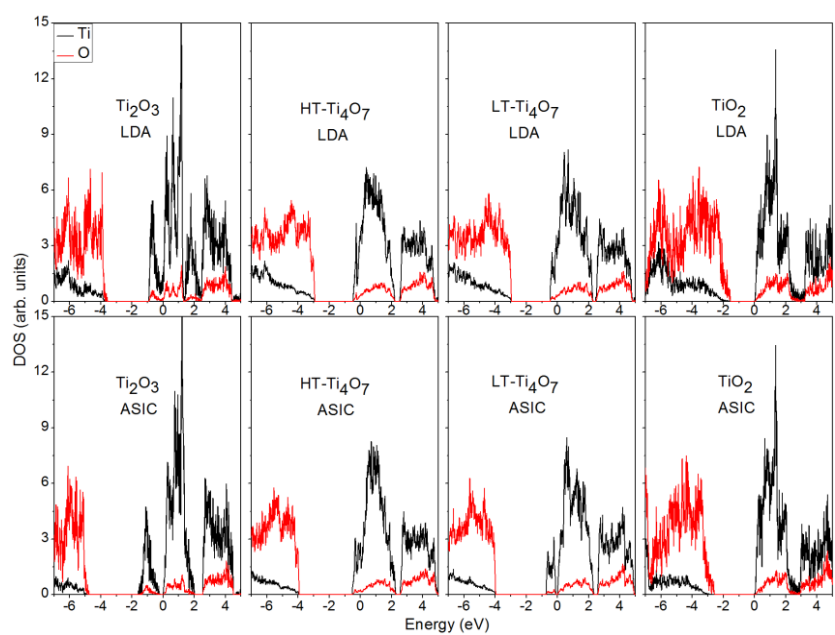

Figure 6. The evolution of non-magnetic density of states as a function of Ti-O stoichiometry. Upper panels: the LDA results. Lower panels: the ASIC results ($\alpha = 0.5$). The black and red curves represent the projected density of states on Ti and O, respectively. For the metallic systems the Fermi level (energy zero) position is uniquely defined, and for the semiconducting systems we align it to the CBM.

As we have mentioned earlier, it is of considerable interest to ascertain if the ASIC approach can be applied to describe electronic properties of an oxide with a mixture of different Ti oxidation states, for example, such that are found during resistive switching. To check this, we apply the same method as we used for $Ti_4O_7$ to rutile $TiO_2$ and $Ti_2O_3$, in which titanium and oxygen are subject to different chemical environment as a result of different stoichiometry. Figure 6 shows comparison of the densities of states for three different stoichiometries. The conduction band is dominated by Ti-3d orbitals, while the valence band is dominated by O-2p orbitals for all phases. All these Ti-O crystals have distorted octahedral crystal fields for the Ti sites. As a result, there is always an $e_g$-$t_{2g}$ splitting of the conduction band across all these crystals. For Ti$_4$O$_7$ and Ti$_2$O$_3$, the deficiency of oxygen results in an additional splitting of the t$_{2g}$ orbitals near the Fermi level.

The conventional LDA functional yields a semiconducting state for TiO$_2$ only, while all other Ti-O phases are calculated to be metallic (finite DOS at Fermi level), which is in contrast to experiments. ASIC on the other hand corrects spurious electronic self-interactions and facilitates charge localization. This results in an increase of band gap for TiO$_2$ (from 1.56 eV at LDA to 2.62 eV at $\alpha = 0.5$). We note that these values are slightly smaller than those predicted using plane-wave basis set implementation (1.88 eV at LDA[40] and 2.9 eV at the VPSIC[40] at $\alpha = 0.5$). Within the ASIC approach, a finite gap is opened for Ti$_2$O$_3$ and LT-Ti$_4$O$_7$, but HT-Ti$_4$O$_7$ remains metallic. It is interesting to note that by applying the ASIC method even the non-polarized picture qualitatively agrees with experiments, i.e., while HT-Ti$_4$O$_7$ is calculated to be a metal, all other oxides are calculated to have a finite gap (Fig. 6). When spin-polarization is taken into account there is no qualitative change to this picture, as will be shown in the following.

TABLE III. The band gaps (eV) of different Ti-O phases calculated by ASIC and HSE06.

|  | TiO$_2$ | LT-Ti$_4$O$_7$ (AF3) | Ti$_2$O$_3$ |
|---|---|---|---|
| ASIC ($\alpha = 0.5$) | 2.62 | 0.94 | 0.23 |
| HSE06 | 3.24 | 1.47 | 0.57 |

When spin polarization is considered, for LT-Ti$_4$O$_7$ the Ti$^{+3}$ ions significantly localize electrons and induce local magnetic moment, which increases band gap from 0.1 eV to 0.94 eV (in LT-AF3 ground state, see Table II). For HT-Ti$_4$O$_7$ there are no such Ti$^{+3}$ ions and it remains metallic (in HT-FM ground state). Next, we compare ASIC functional results to HSE06 hybrid functional in describing electronic properties of these Ti-O phases (Table III). Both methods yield the same ordering and similar energy differences for band gap change vs. stoichiometry, while HSE06 predicts a few tenths of eV wider gap for all phases. The predicted band gaps for LT-Ti$_4$O$_7$ by both methods are larger than the experimentally reported values. Note however that in this case the experimentally measured band gaps might not correspond to the fundamental gap; rather, they may reflect the energy required to flip two spins[22].

Finally, to ascertain that the chosen value of ASIC correction strength α is appropriate for different Ti-O stoichiometries, we show the predicted band gaps of all gapped Ti-O phases considered in this work, as a function of α. All these materials exhibit a nearly linear relationship between band gap and α in the regions of α with non-zero band gaps as shown in Figure 7. This is consistent with the previous ASIC benchmark work.[25] The appropriate value of α is in principle expected to vary from one material to another, considering the fact that α is related to the screening properties of a given material. On the other hand, we find that a single value of $\alpha = 0.5$ can properly describe all the Ti-O phases shown here, at least as far as the electronic conduction properties are concerned, with band gaps reasonably well reproduced. Depending on the specific Ti-O phases, the HSE06 functional behaves (in terms of calculated band gap) like ASIC from $\alpha = 0.6$ to $\alpha = 0.7$. Note however that the computational cost of an ASIC calculation is much lower than a corresponding HSE06 calculation. For example, for calculating the rutile $TiO_2$ electronic structure at a similar accuracy ASIC costs less than one tenth of HSE06 does.

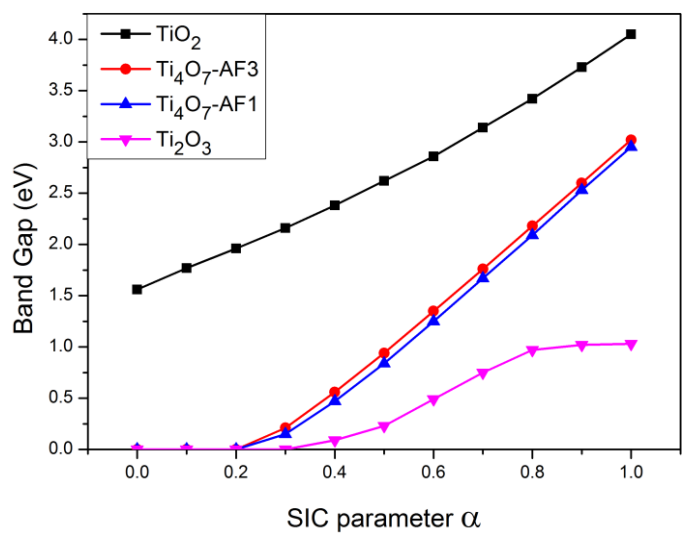

Figure 7. Predicted band gap vs. empirical parameter α for LT-$Ti_4O_7$, $TiO_2$ and $Ti_2O_3$.

## 4. Conclusions

In summary, we have systematically used the LDA-ASIC approach to study the $Ti_4O_7$ Magnéli phase, as well as $Ti_2O_3$ and rutile $TiO_2$. By searching throughout the possible spin configurations we predicted a new antiferromagnetic semiconducting ground state (AF3) and show that there are two competing semiconducting states (AF1 and AF3) with similar energies. We propose a Heisenberg model to explain the lower energy of the AF3 state compared with the AF1 state. This result is also confirmed using a hybrid functional (HSE06) approach. The HT phase on the other hand is found to be metallic. For LT- $Ti_4O_7$ the competition of AF1 and AF3 states is tested to be robust in a wide range of isotropic strain, while for HT phase there appears to be a magnetic phase transition at compressive strain. Our results show that both ASIC and HSE06 functional are capable of describing electronic properties of $Ti_4O_7$, $Ti_2O_3$ and rutile $TiO_2$, with ASIC requiring only a fraction of the computational cost of an HSE06 calculation. This paves the way to modeling of resistive switching of Ti-O based heterostructures within the framework of density functional theory.

## ACKNOWLEDGMENTS

Work at Argonne National Laboratory was supported by the U.S. Department of Energy, Office of Science, Office of Basic Energy Sciences, under contract DE-AC02-06CH11357. I.R. acknowledges financial support from the EU project ACMOL (FP7-FET GA618082).We gratefully acknowledge the computing resources provided on Blues and Fusion, high-performance computing clusters operated by the Laboratory Computing Resource Center at Argonne National Laboratory.